\documentclass[]{raa}
\usepackage[utf8]{inputenc}
\usepackage{graphicx,times}             
\usepackage{natbib}
\usepackage{amssymb,amsmath}
\bibpunct{(}{)}{;}{a}{}{,}
\usepackage[OT1]{fontenc}
\usepackage[]{hyperref}
\hypersetup{colorlinks = true, linkcolor = blue, anchorcolor = red, citecolor = blue, filecolor = red, urlcolor = blue}

\newcommand{\s}{PKS 1749+096}
\newcommand{\sr}{PKS 1749+096~}

\begin{document}

\title{Resolving the inner jet of \sr with super-resolution VLBA images at 7 mm}

   \volnopage{Vol.0 (20xx) No.0, 000--000}      
   \setcounter{page}{1}          

   \author{
          Lang Cui
            \inst{1,2}
     \and Ru-Sen Lu
            \inst{3,2,4}\thanks{Corresponding author}
     \and Wei Yu
            \inst{3}
     \and Jun Liu
            \inst{4}
     \and V{\'i}ctor M. Pati{\~n}o-{\'A}lvarez
           \inst{5,4}
     \and Qi Yuan
            \inst{1,6}
    }


   \institute{
             Xinjiang Astronomical Observatory, Chinese Academy of Sciences, Urumqi 830011, China; {\it cuilang@xao.ac.cn}\\
        \and
             Key Laboratory of Radio Astronomy, Chinese Academy of Sciences, Nanjing 210008, China\\
        \and
             Shanghai Astronomical Observatory, Chinese Academy of Sciences, Shanghai 200030, China; {\it rslu@shao.ac.cn}\\
        \and
             Max Planck Institute for Radio Astronomy, Auf dem h\"ugel 69, Bonn 53121, Germany\\
        \and
            Instituto Nacional de Astrof\'isica, \'Optica y Electr\'onica,  Luis Enrique Erro $\#$ 1, Tonantzintla, Puebla 72840, M\'exico\\
        \and
            University of Chinese Academy of Sciences, Beijing 100049, China\\
\vs\no
   {\small Received~~20xx~~month day; accepted~~20xx~~month day}}

\abstract{High resolution imaging of inner jets in Active Galactic Nuclei (AGNs) with VLBI at millimeter wavelengths provides deep insight into the launching and collimation mechanisms of relativistic jets. The BL Lac object, \s, shows a core-dominated jet pointing toward the northeast on parsec-scales revealed by various VLBI observations. In order to investigate the jet kinematics, in particular, the orientation of the inner jet on the smallest accessible scales and the basic physical conditions of the core, in this work we adopted a super-resolution technique, the Bi-Spectrum Maximum Entropy Method (BSMEM), to reanalyze VLBI images based on the Very Long Baseline Array (VLBA) observations of \sr within the VLBA-BU-BLAZAR 7\,mm monitoring program. These observations include a total of 105 epochs covering the period from 2009 to 2019. We found that the stacked image of the {inner jet} is limb-brightened with
an apparent opening angle of $50\fdg0\pm8\fdg0$ and $42\fdg0\pm6\fdg0$ at the distance of 0.2 and 0.3\,mas (0.9 and 1.4\,pc) from the core, corresponding to an intrinsic jet opening angle of $5\fdg2\pm1\fdg0$ and $4\fdg3\pm0\fdg7$, respectively. In addition, our images show a clear jet position angle swing in \sr within the last ten years. We discuss the possible implications of jet limb brightening and the connection of the position angle with jet peak flux density and gamma-ray brightness.
}

   \authorrunning{L. Cui et al. }
   \titlerunning{Resolving the inner jet of \s}

   \maketitle

\section{Introduction}
\label{sect:intro}

High resolution observations of jets in Active Galactic Nuclei (AGNs) with Very Long Baseline Interferometry (VLBI) show that some objects have regular or irregular swings in the position angle of the innermost structure~\citep[i.e., jet wobbling,][]{Agudo+etal+2007, Lu+etal+2012}. The physical origin of this phenomena is currently still not well understood. Possible mechanisms for the jet wobbling include accretion disk precession, orbital motion of accretion systems (binary black hole), and jet instabilities~\citep[e.g.,][]{Agudo+2009}.
In some sources, a correlation between the variations of the inner jet position angle and of the flux density
in radio, X-ray, or gamma-ray is observed~\citep[e.g.,][and references therein]{Rani+etal+2014}, suggesting a connection between the inner jet morphology and the corresponding emitting regions.

At a redshift of 0.322~\citep{Stickel+etal+1988}, the ultra-luminous BL Lac object \sr showed strong variability from radio to X-ray regime. In 2016 July, unprecedented bright flaring activity in very high energy gamma-ray emission, together with X-ray and optical flares were detected~\citep{Becerra Gonzalez+etal+2016, Ciprini+etal+2016, Mirzoyan+2016, Balonek+etal+2016}. The radio morphology of \sr is dominated by its mas-scale structure. Multi-epoch VLBI observations show that position angle swing exists in its jet~\citep{Lu+etal+2012}, which is more pronounced towards the core.

To investigate the structure of the innermost region of the jet and its possible variations, here we report the results of a detailed VLBA imaging study of \sr at 7\,mm with unprecedented resolution. Through this paper, we adopt the following parameters: the luminosity distance to \sr is $D_{L}$=1674\,Mpc and 1\,mas of angular separation corresponds to 4.64\,pc, with cosmological parameters $\Omega_M$=0.27, $\Omega_{\kappa}$=0.73, and $H_0$=71\,km$s^{-1}$.

\section{Archival Data and VLBI Images}
\label{sect:data}

For our study, we made use of data from the VLBA-BU-BLAZAR 7\,mm monitoring program~\citep[e.g.,][]{Jorstad+etal+2017}, spanning a time interval from 2009 April to 2019 October, during which 105 epochs of observations were performed with an approximately monthly cadence. The typical resolution of the CLEAN images from these observations is $\sim$0.3 mas and $\sim$0.15\,mas in the north-south and east-west direction, respectively.

Recently, various super-resolution imaging techniques have been developed and demonstrated for black hole shadow imaging in Sgr\,A* and M87 and reconstruct of polarization structures with VLBI polarimetry observations~\citep[e.g.,][and references therein]{Coughlan+Gabuzda+2016, Chael+etal+2016, Akiyama+etal+2017a, Akiyama+etal+2017b, Kuramochi+etal+2018}.
These methods have been demonstrated that better resolution (by a factor of $\gtrsim$ 2) than that is normally achieved with the traditional CLEAN methods can be obtained. With these techniques, the obtained resolution at 7\,mm is comparable to that obtained with the traditional CLEAN method with 3mm-VLBI of a similar array.

\begin{figure}
\centering
\includegraphics[width=0.45\textwidth,clip]{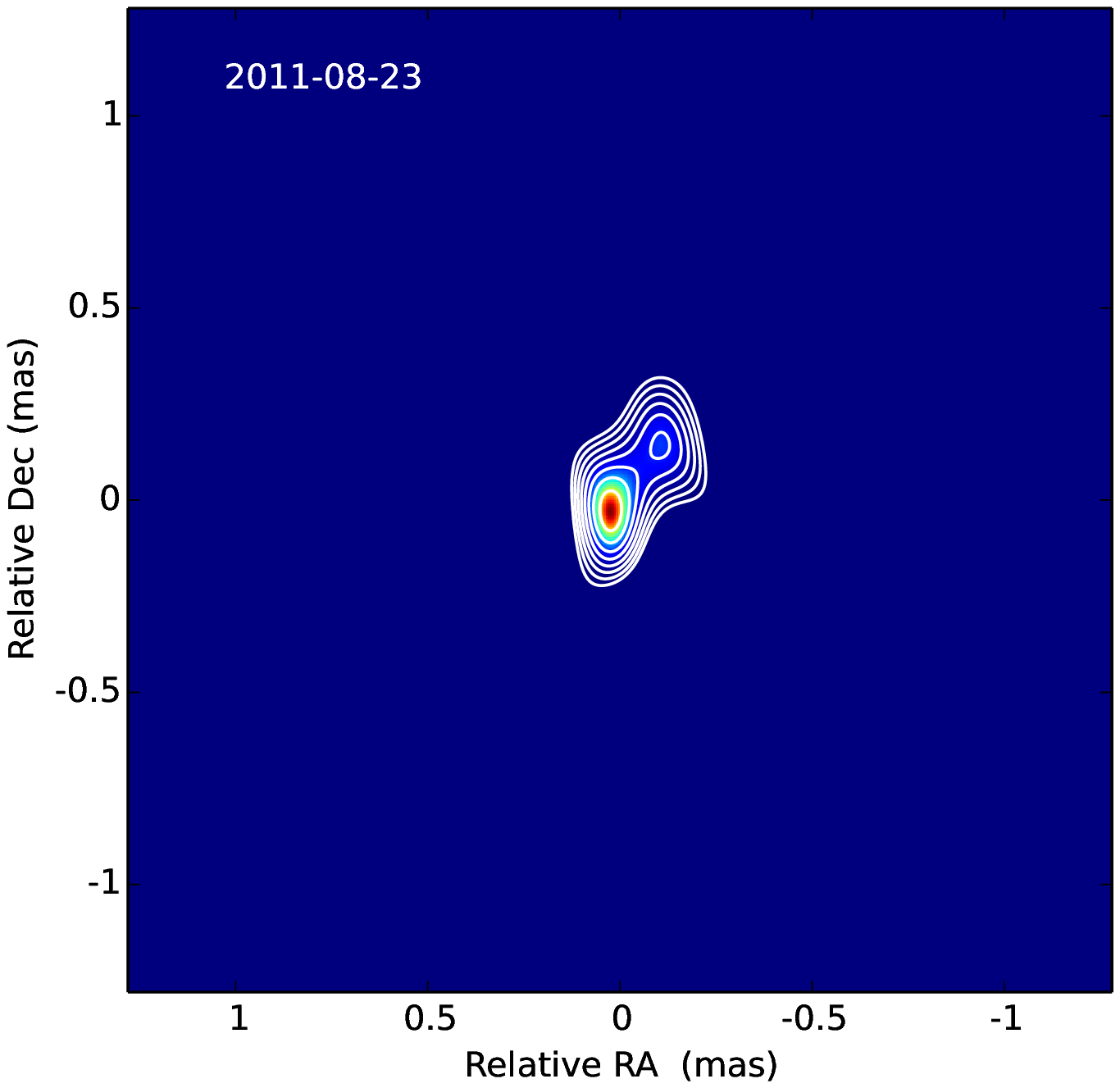}
\includegraphics[width=0.45\textwidth,clip]{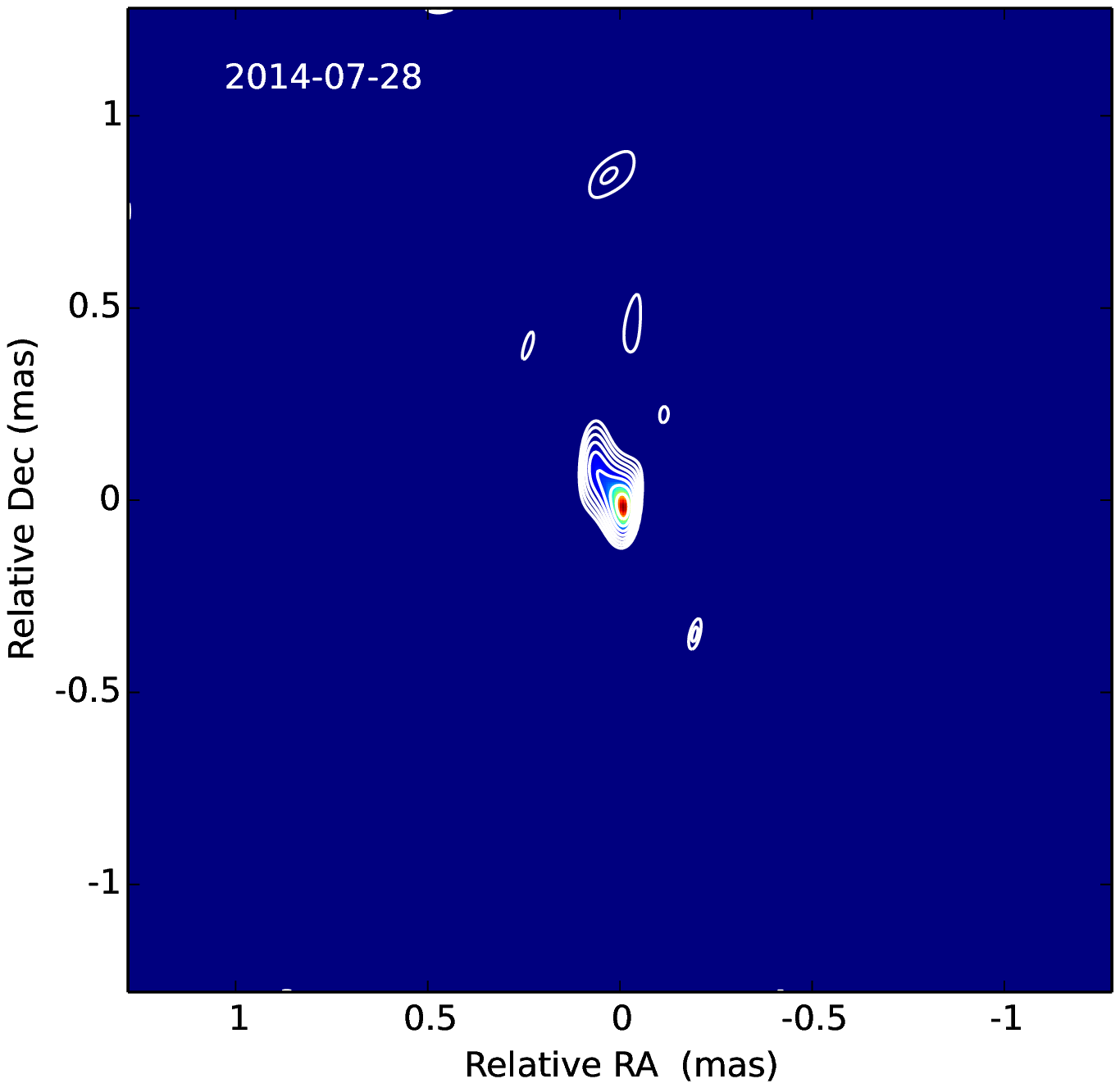}
 \caption{BSMEM images of \sr in 2011 August and 2014 July. For each image, contours start at 0.5\,\% of the peak brightness in steps of two.}
\label{fig:mem}%
\end{figure}

In order to investigate the fine-scale structure in the core region, where the jet position angle swing seems to be more pronounced, we adopted a super-resolution interferometric imaging technique, the Bi-Spectrum Maximum Entropy Method \citep[BSMEM,][]{Buscher+1994} and re-imaged the inner jet. Applications of this method to simulated and observed data of M87 show that it produces high-resolution images with reliable structures~\citep{Lu+etal+2014, Kim+etal+2016}. As examples, we show the BSMEM reconstructed images of \sr in 2011 August and 2014 July in Figure~\ref{fig:mem}.

\section{Results \& Discussion}
\label{sect:results}

\begin{figure*}
\centering
\includegraphics[width=0.4\textwidth,clip]{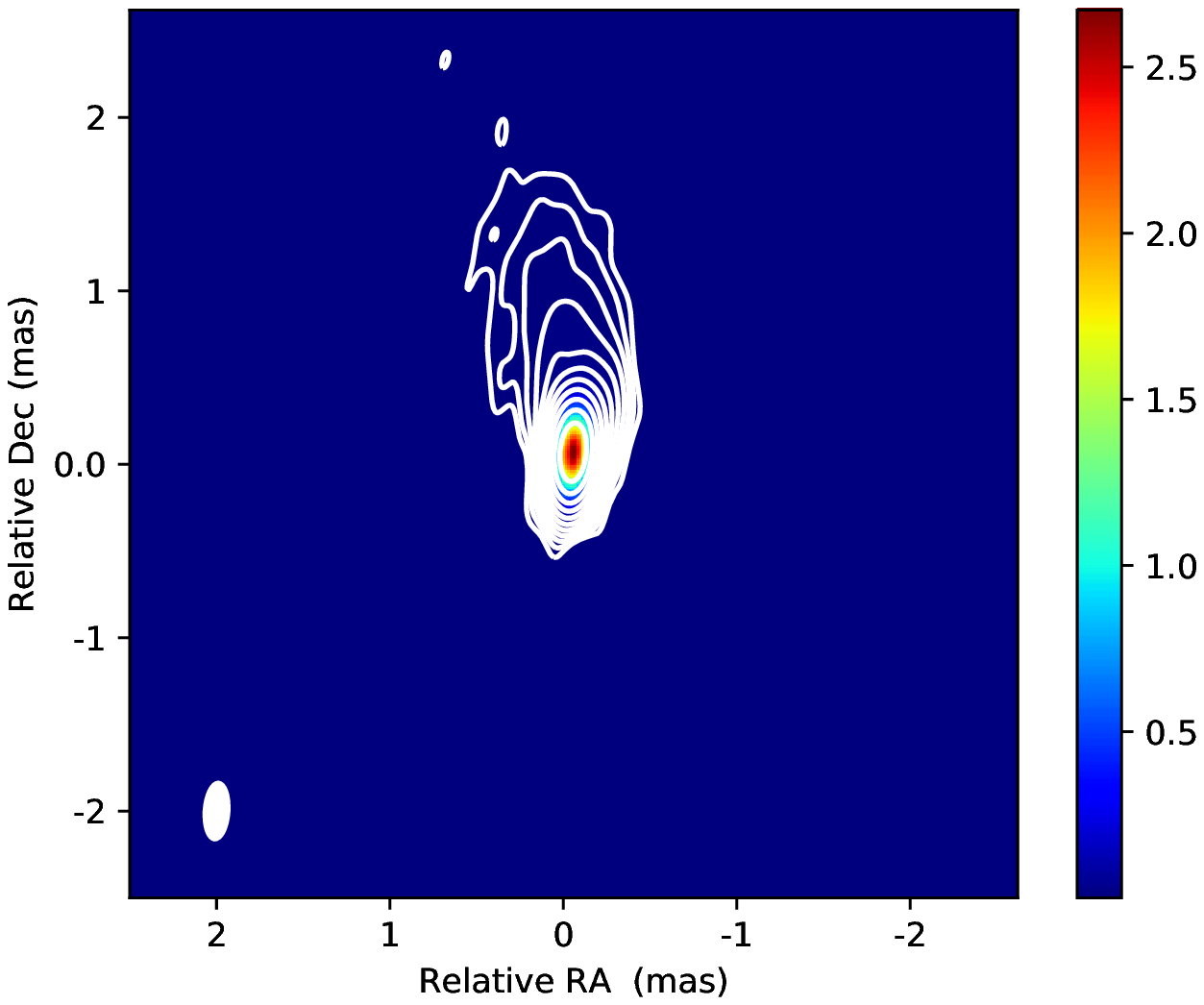}
\includegraphics[width=0.556\textwidth,clip]{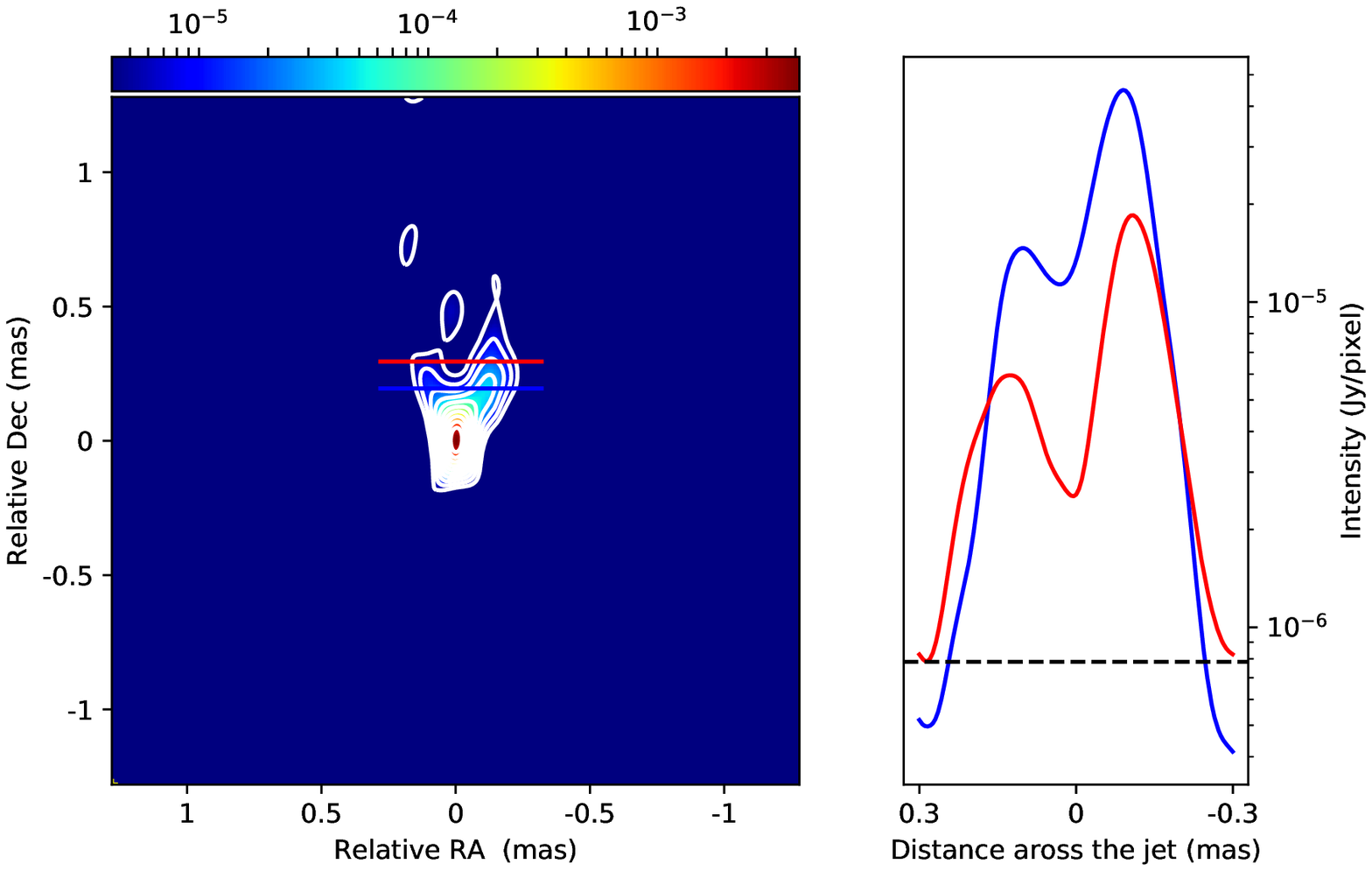}
 \caption{(Left) Stacked CLEAN image of \s. Before stacking, each epoch is restored with a common restoring beam of 0.33$\times$0.14 at -4\fdg0, corresponding to the median uniformly weighted beam of all epochs. Contours start at 0.05\,\% of the peak in steps of two. The color scale is given in Jy/beam. (Middle) Stacked BSMEM image of \sr at 7\,mm. Contours start at 0.1\,\% of the peak in steps of two. The color scale is given in Jy/pixel, with a pixel size of 5$\mu$as. The blue and red lines denote the position of the slices in the right panel. (Right) The transverse jet intensity profiles measured along the blue and red lines in the middle panel at core separation of 0.2 and 0.3\,mas, respectively. The dashed line marks $5\sigma$ off-source image RMS noise level.}
\label{fig:stacked}%
\end{figure*}

To show the overall extent of the jet at 7\,mm, we produced a ``stacked'' uniformly weighted image (Figure~\ref{fig:stacked}, left). We created the stacked image by first restoring the image at each epoch with an identical CLEAN beam whose dimensions corresponded to the median uniformly weighted beam of all epochs. We then shifted the convolved image at each epoch to align the fitted positions of the image peak, and then combined them all with equal weight to produce an averaged (stacked) image. At 7\,mm, the overall VLBI-scale jet is dominated by the bright core, with an extended emission pointing toward north-east, consistent with the jet morphology seen at other epochs~\citep{Lu+etal+2012}.

\subsection{Transverse Jet Structure and its Implications}
The high-resolution BSMEM images revealed detailed structure of the inner jet (within $\sim$0.5\,mas of the core), which cannot be clearly seen in normal CLEAN images. A visual inspection of the position angle of the innermost portion of the jet in these images indicates a variation of the position angle with time. This suggests that the jet at a given epoch may only occupy a portion of the whole underlying jet. If this is the case, we can stack the BSMEM images using all the available epochs to reconstruct the whole jet. \citet{Pushkarev+etal+2017} found that a true jet geometry appears only after stacking epochs over several years for a considerable fraction of their studied sample of AGNs. Interestingly, the transverse jet structure of the stacked BSMEM image of \sr is transversely resolved and shows a limb-brightened morphology (Figure~\ref{fig:stacked}, middle). The limb brightening is most pronounced in the inner part of the jet ($<$0.4\,mas), while the jet farther out is resolved out and invisible.

By determining the transverse jet width, we can measure the apparent opening angle of the underlying jet with the caveat that such a stacked opening angle is different from a single epoch jet opening angle. We found that the apparent jet opening angle is $50\fdg0\pm8\fdg0$ at the core distance of 0.2\,mas (0.9\,pc) and $42\fdg0\pm6\fdg0$ at the core distance of 0.3\,mas (1.4\,pc) (Figure~\ref{fig:stacked}, right), suggesting that the jet still undergoes collimation on these scales. We note that \citet{Finke+2019} adopted a smaller jet-opening angle of 16\fdg8 for \sr in a recent study of physical properties for a sample of blazar radio jets, but this result was obtained based on a single-epoch 15\,GHz observation with lower resolution~\citep{2009A&A...507L..33P}.
With the measured apparent jet opening angle ($\alpha_{\rm app}$), the intrinsic jet opening angle ($\alpha_{\rm int}$) can be calculated according to the relation of tan($\alpha_{\rm int}/2$)=tan($\alpha_{\rm app}/2$)sin$\theta$~\citep{Pushkarev+etal+2017}, where $\theta$ is the jet viewing angle. Here, we find the intrinsic jet opening angle is $5\fdg2\pm1\fdg0$ at the core distance of 0.2\,mas and $4\fdg3\pm0\fdg7$ at the core distance of 0.3\,mas by adopting a viewing angle of $\theta=5\fdg6$.

Limb brightening in jet structures has been detected in a couple of nearby objects, e.g., M87~\citep[][]{Kim+etal+2018}, Cygnus A~\citep[][]{Boccardi+etal+2016}, 3C\,84~\citep{2018NatAs...2..472G}. Such a stratified structure can be understood as the result of a fast internal spine and a slower external sheath in the jet~\citep[e.g.,][]{2005A&A...432..401G}. The detection of limb brightening close to the central black hole in these sources indicates that the velocity structure originates from the jet launching site with the central spine component possibly being associated with the black-hole-powered jet whereas the sheath component anchored to the accretion disk. For a black hole mass of $10^{8.34}\,M_\odot$~\citep{Zhou+Cao+2009}, a core distance of 0.2--0.3\,mas corresponds to 4.5--6.8$\times10^5$ Schwarzschild Radii (deprojected, for a viewing angle of $\theta=5.6^{\circ}$) in \s. In this context, it may seem surprising to detect a limb-brightened underlying structure originated from the jet base. On the other hand, a velocity structure could also originate from the interaction of the jet with the ambient medium driven by Kelvin-Helmholtz instabilities of the jet~\citep[e.g.,][]{2008A&A...488..795R}. In this scenario, the ambient/jet density contrast is a key parameter determining the instability evolution and the entrainment properties of the external medium. Future higher resolution observations would help distinguish between these two scenarios.

\subsection{Wobbling of the Inner Jet}

In some jetted sources, correlated variations between the inner jet position angle and flux density in radio, X-rays, and gamma-rays have been observed~\citep[e.g.,][and references therein]{Liu+etal+2012,Rani+etal+2014}. \citet{Rani+etal+2014} found a significant correlation between the Fermi-LAT (Large Area Telescope) gamma-ray flux variations and the position angle variations in the VLBI jet of S5\,0716+714. To explain this correlation, the authors proposed a scenario where a moving shock propagating down a (bending) relativistic jet causes significantly increased emission, when the gamma-ray emission and orientation of the jet flow share the same boosting cone, a correlation between the two is expected.

\begin{figure}[h!]
\centering
\includegraphics[width=1.0\textwidth,clip]{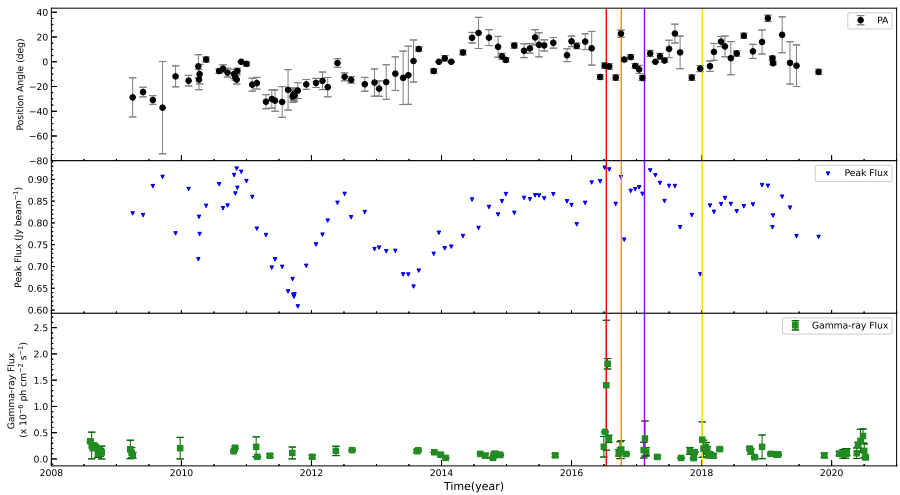}
 \caption{Time evolution of the position angle (PA) of the innermost jet within 0.3\,mas of the core (top) and peak flux density (middle), and weekly averaged gamma-ray flux light curve in the energy range from 0.1 to 300 GeV. Vertical lines mark the time of the big gamma-ray flare in 2016 July (red line) and three major gamma-ray brightening events after it.}
\label{fig:fg3}%
\end{figure}

In Figure~\ref{fig:fg3} (top panel), we show the time evolution of the position angle of the innermost jet. In calculating the position angle, we first determined the jet ridgeline, which is the line connecting the maxima of the jet brightness profiles measured transversally to the local jet direction. Then the mean position angle of evenly spaced points on the jet ridgeline within 0.3\,mas of the image peak was used to represent the inner jet position angle. The majority of the position angles range between -30$^\circ$ and 20$^\circ$. Over the time scale covered by the observations presented here, it is clear that no regular or periodic position angle changes exist. Rather, it seems that the inner jet shows predominantly slower changes in its direction but with occasionally erratic features (e.g., the rapid swing during 2011--2012 and 2016). \citet{Lister+etal+2013} studied the jet orientation with VLBA monitoring observations at 2\,cm and they showed that \sr is one of their sample of sources showing oscillatory behavior in the inner jet position angle with a fitted period of 12 years. Figure~\ref{fig:fg3} (middle panel) shows the time evolution of the jet peak flux density. In order to investigate the possible correlation between the inner jet position angle and the peak flux density, we calculated the Kendall's correlation coefficient ($\tau$) and found $\tau$ is 0.26 with a p-value of $7.2\times10^{-5}$, indicating a weak correlation between them.

Figure~\ref{fig:fg3} (bottom panel) shows the Fermi-LAT gamma-ray light curve of \s. The Fermi-LAT gamma-ray data were used in the energy range from 0.1 to 300 GeV, and processed using the Fermitools version 1.0.2. We analyzed a 15-degree-radius region of interest around the position of 1749+096, and added all sources within, which were extracted from the 4FGL catalogue \citep{Abdollahi+etal+2020}. The normalization factors and spectral parameters for sources within 5 degrees were kept free; while on sources farther than 5 degrees only their normalization parameter is free to vary. We also applied up-to-date diffuse and isotropic background models along with the current set of instrument response functions. We binned the data in 7-days intervals to increase the signal to noise ratio without compromising temporal resolution. Only the data points with a Test Statistic (TS) value greater than 25 (significant detection) were used.

Since \sr was only intermittently detectable by the Fermi-LAT and weakly variable most of the time during the last ten years, a correlation analysis for the gamma-ray flux and inner jet position angle yields a Kendall's $\tau$ of -0.11 with a p-value of 0.02. This indicates that there are no strong evidence for a correlation between the orientation of the inner jet and gamma-ray flux, which suggests that the gamma-ray emission in \sr is not driven by orientation-dependent effects, consistent with the recent findings for radio galaxies and blazars~\citep[][and references therein]{2019A&A...627A.148A}. We note that \sr showed unprecedented flares in 2016 July in multiple wavebands, including optical, X-rays, and gamma-rays ~\citep{Balonek+etal+2016, Ciprini+etal+2016, Mirzoyan+2016, Schussler+etal+2017} and the source significantly brightened a few (three) times in gamma-rays after the major flare in 2016 July. Figure~\ref{fig:fg3} marks the time of these events with accompanying jet position angle measurements. Interestingly, these events seemed to occur when the inner jet was along a similar position angle ($\sim$-10$^\circ$), although no significant gamma-ray flux increase was seen for similar jet position angles prior to 2016. Future observations with more gamma-ray flux measurements would allow further investigation of their possible connection to the inner jet orientation.

\section{Summary}
\label{sect:summary}
In this work, we presented results of a detailed VLBA imaging study of \sr at 7\,mm with unprecedented resolution based on a Maximum Entropy Method technique. We found that the stacked image of the inner jet has a limb-brightened structure with apparent opening angles of $50\fdg0\pm8\fdg0$ and $42\fdg0\pm6\fdg0$ at the distance of 0.2 and 0.3\,mas (0.9 and 1.4\,pc) from the core, corresponding to an intrinsic jet opening angle of $5\fdg2\pm1\fdg0$ and $4\fdg3\pm0\fdg7$, respectively. The double peaked transverse jet profile is suggestive of an underlying ``spine-sheath''  structure possibly related to the jet launching process or the interaction of the jet with its ambient medium. We also found that the jet position angle in \sr shows a clear swing phenomenon within the last 10 years, which weakly correlates with the peak brightness. The lack of correlation between the inner jet position angle and gamma-ray flux suggests that the gamma-ray emission in \sr is not sensitive to orientation-dependent effects.

\begin{acknowledgements}
We thank the anonymous referee for valuable comments and suggestions. This study made use of 43 GHz VLBA data from the VLBA-BU Blazar Monitoring Program (VLBA-BU-BLAZAR; http://www.bu.edu/blazars/VLBAproject.html), funded by
NASA through the Fermi Guest Investigator Program. The VLBA is an instrument of
the National Radio Astronomy Observatory. The National Radio Astronomy Observatory is a facility of the National Science Foundation operated by Associated Universities, Inc.

L. C. thanks support from the National Key R\&D Program of China (grant No. 2018YFA0404602) and the CAS 'Light of West China' Program (grant No. 2018-XBQNXZ-B-021). R.-S. Lu and V. M. Pati{\~n}o-{\'A}lvarez are supported by the Max Planck Partner Groups at SHAO and INAOE, respectively. R.-S. Lu acknowledges the support by the Key Program of the National Natural Science Foundation of China (NSFC grant No. 11933007) and the Research Program of Fundamental and Frontier Sciences, CAS (grant No. ZDBS-LY-SLH011). This work is also supported in part by NSFC under grant No. 61931002 \& U1731103 and the Youth Innovation Promotion Association of the CAS under grant No. 2017084.

\end{acknowledgements}

\label{lastpage}
\end{document}